\documentclass{elsarticle}

\usepackage{natbib}

\usepackage{graphicx}
\usepackage{subfigure}

\usepackage{amssymb}

\begin{document}

\begin{frontmatter}

\title{Study of the variability of Blazars gamma-ray emission}

\author{T. Sbarrato$^a$\footnote{E-mail: t.sbarrato@campus.unimib.it, tullia.sbarrato@brera.inaf.it}, 
L. Foschini$^b$, G. Ghisellini$^b$, F. Tavecchio$^b$}
\address{$^a$Department of Physics G. Occhialini, University of Milano-Bicocca,
Piazza della Scienza 3, 20126 Milano, Italy\\
$^b$INAF - Osservatorio Astronomico di Brera, via E.\ Bianchi 46, 23807, Merate (LC), Italy}

\begin{abstract}
The $\gamma$-ray emission of blazar jets shows a pronounced variability and this feature provides limits to the size and to the speed of the emitting region.
We study the $\gamma$-ray variability of bright blazars using data from the first 18 months of activity of the Large Area Telescope on the \textit{Fermi Gamma-Ray Space Telescope}.
From the daily light-curves of the blazars characterized by a remarkable activity, we firstly determine the minimum variability time-scale, giving an upper limit for the size of the emitting region of the sources, assumed to be spheroidal blobs in relativistic motion. 
These regions must be smaller than $\sim10^{-3}$ parsec.
Another interesting time-scale is the duration of the outbursts. 
We conclude that they cannot correspond to radiation produced by a single blob moving relativistically along the jet, but they
are either the signature of emission from a standing shock extracting energy from a modulated jet, or the superposition of a number of flares occurring on a shorter time-scale.
We also derive lower limits on the bulk Lorentz factor needed to make the emitting region transparent for gamma-rays interacting through photon-photon collisions.

\end{abstract}


\end{frontmatter}

\parindent=0.5 cm

\section{Introduction}

Blazars are active galactic nuclei (AGN) with a relativistic jet pointing at a small angle from our line of sight.
This feature allows to observe strong relativistic effects in blazars emissions, such as the shortening of time intervals, and hence rapid variability.
Variability at all frequencies is indeed a defining characteristics of  this class of sources
(see for example Ulrich et al.~1997).
The study of variability at high energies is particularly interesting, because most of the bolometric luminosity of these objects is emitted at $\gamma$-rays (MeV-GeV-TeV).

The EGRET experiment onboard the \textit{Compton Gamma-Ray Observatory} already detected rapid variability with time-scales of few hours.
As an example, EGRET measured in January 1996 a flux increase of a factor of $2.6$ in $\sim8$ hours in the emission of the well-known FSRQ 3C~279 (Wehrle et al.~1998).
In June 1995, EGRET observations of PKS~1622-297 led to the detection of a huge flare, with a flux increase of a factor of $3.6$ occurred in less than $7.1$ hours, with a corresponding doubling time of less than $3.8$ hours (Mattox et al.\ 1997).
The early results obtained from the Large Area Telescope (LAT) onboard the \textit{Fermi Gamma-Ray Space Telescope} (\textit{Fermi}) show rapid variability of few hours (see Foschini et al.\ 2010, Tavecchio et al.\ 2010, Abdo et al.\ 2010).

Such a rapid variability is useful to constrain the size $R$ of the emitting region of blazar jets.
The latter is linked to the minimum variability time-scale ($t_{min}$) by the relationship $R < c\, t_{min} \delta\,(1+z)^{-1}$, where $\delta$ is the relativistic Doppler factor ($\delta=[\Gamma(1-\beta\cos\theta)]^{-1}$, where $\beta c$ is the emitting region velocity, $\theta$ the angle of its motion with respect to our line of sight and $\Gamma$ the bulk Lorentz factor).
Assuming the emitting region to be a sort of spherical blob that increases its size following the jet aperture, we can obtain an estimate of its distance from the central engine.
Calculating the size of the emitting region allows to evaluate at which distance from the central black hole the jet originates.
This kind of result would give important constraints on jet modeling, because understanding how the jet originates near the central black hole is nowadays an interesting challenge of blazar paradigm (for a closer examination about blazar detailed structure see Begelman, Blandford and Rees, 1984).

An estimate of the mentioned parameters ($R,\delta$) results in general agreement with the observed time-scales of few hours range.
Some results obtained with the High Energy Stereoscopic System (HESS), however, give evidence of very fast variability at $\gamma$-rays in the TeV BL Lac PKS 2155-304 (Aharonian et al.~2007, for multiwavelenght follow-up Foschini et al.~2007 and Aharonian et al.~2009).
During an exceptional outburst in July 2006, this source showed an activity characterized by a doubling time-scale of $\sim200$~s.
Another example of very fast variability has been detected by the Major Atmospheric Gamma-ray Imaging Cerenkov Telescope (MAGIC) during the remarkable activity of the TeV BL Lac Mrk 501 in June and July 2005 (Albert et al.\ 2007).
In this case the variability time-scale derived during the most active nights was of $\sim5$ min.
With the current values of the parameters mentioned above, results like those obtained for PKS 2155-304 and Mrk 501 are about one order-of-magnitude less than what one expects, i.e.\ $t_{min} \sim R_{s}/c$ (Begelman et al.\ 2008).
A solution of this problem could consist in increasing the value of the Doppler factor, i.e.\ the relativistic beaming of the jet.

However, very large values of the bulk Lorentz factor could be possible for BL Lacs (Begelman et al., 2008), but are problematic in FSRQs.
This is because, in FSRQs, the high energy emission is thought to be produced through external Compton (EC) mechanism, i.e.\ inverse Compton on a population of seed photons external to the jet.
The energy density of these external photons is enhanced, in the comoving frame, by $\sim\Gamma^2$.
A too large value of $\Gamma$ makes the $\gamma -\gamma\rightarrow e^{\pm}$ process important, and $\gamma$-rays would be absorbed.
This limit is not present in BL Lacs, due to their lack of emission lines.

The Large Area Telescope (Atwood et al.\ 2009) represents a good instrument to improve studies about blazar variability in the MeV-GeV energy band.
The great advantage of LAT operating in scanning mode is to offer an efficient monitoring of the whole sky, with a complete coverage every three hours, that leads to the possibility of drawing 3 hours-binned light curves, i.e.\ the finest binning achievable in scanning mode.
This feature can eventually allow to deepen the study of minimum variability.
Because of \textit{Fermi}/LAT useful features, our study is focused on the analysis of the $\gamma$-ray variability of bright blazars detected by this telescope during the period from August 4, 2008 to January 25, 2010.

In the present work we use a cosmology with $H_0 = 71\;\textrm{km s}^{-1}\textrm{Mpc}^{-1}$, ${\Omega}_{\Lambda}=0.7$ and ${\Omega}_{M}=0.3$.

\section{Data and analysis}

Data of \textit{Fermi}/LAT  have been downloaded from the publicly accessible web site of HEASARC\footnote{\tt http://fermi.gsfc.nasa.gov/\rm\normalsize}, together with the software of analysis.
We focused on five sources characterized by a pronounced variability, selected from a public list of monitored sources.
According to LAT data policy, a source belongs to this list if its flux exceeds a minimum value of $2\cdot10^{-6}\textrm{ph cm}^{-2}\textrm{s}^{-1}$ ($E>100$~MeV). 
In order to identify the most variable sources, we firstly made an overall analysis of the whole LAT list.
From the data published  we obtained 1 day-binned light curves covering all the time period.
The minimum statistical confidence level accepted for each time bin is $TS=25$, where $TS$ is the test statistic described in Mattox et al.\ (1996). 
This statistical characterization is provided by HEASARC among the data published in the monitored sources list.
Therefore, according to this selection, we generate light curves taking only time bins with $\sigma \sim \sqrt{TS} > 5$.
Because of this statistical selection, each source will have a different number of time bins.
Therefore, we take into account that the less statistically significant light curves might have time intervals without reliable data.

We selected the most variable sources working on the flux deviation from the average flux value.
In order to characterize this deviation, we calculated the $\chi^{2}_{red}$ with respect to a constant flux model centered on the average flux. 
The minimum value accepted is $\chi^{2}_{red}=8$.
We found five sources out of the sample that fulfill this condition: 0235+164, 3C 273, 3C 279, 1510-089, 3C 454.3 (for the specific $\chi^{2}_{red}$ values, see Table \ref{table1}).
We focuss on this smaller group of sources.
\begin{table}
\centering
\begin{tabular}{l l l l l l l}
\hline
Source & $z$ & $d_L$ [cm] & $F_x$ [$\mu$Jy] & ${\alpha}_x$ & $N_{points}$ & $\chi^2_{red}$\\
\hline
0235+164 & 0.940 & $1.88\cdot10^{28}$ & 0.34 & 0.44 & 138 & 11.61\\
3C 273 & 0.158 & $2.33\cdot10^{27}$ & 67.77 & 0.56 & 213 & 16.96\\
3C 279 & 0.536 & $9.51\cdot10^{27}$ & 0.98 & 0.76 & 266 & 22.80\\
1510-089 & 0.360 & $5.93\cdot10^{27}$ & 0.44 & 0.40 & 263 & 92.36\\
3C 454.3 & 0.859 & $1.69\cdot10^{28}$ & 2.97 & 0.70 & 404 & 18.41\\
\hline
\end{tabular}
\caption{\small The five sources out of the overall sample that we analysed in details.
$z$, $d_L$,  $F_x$ and ${\alpha}_x$ are the parameter used to calculate the $\delta$ factor (data from Ghisellini et al. 2010).
$N_{points}$ is the number of statistically significant points of the light curves.
$\chi^2_{red}$ represents the deviation of the light curves with respect to a constant flux model centered on the average flux ($N_{points}-1$ is the number of degrees of freedom involved in each calculation).}
\label{table1}
\end{table}

\begin{figure}
\centering
\subfigure[]
   {\includegraphics*[angle=270, scale=0.25, clip, trim=0 0 0 0]{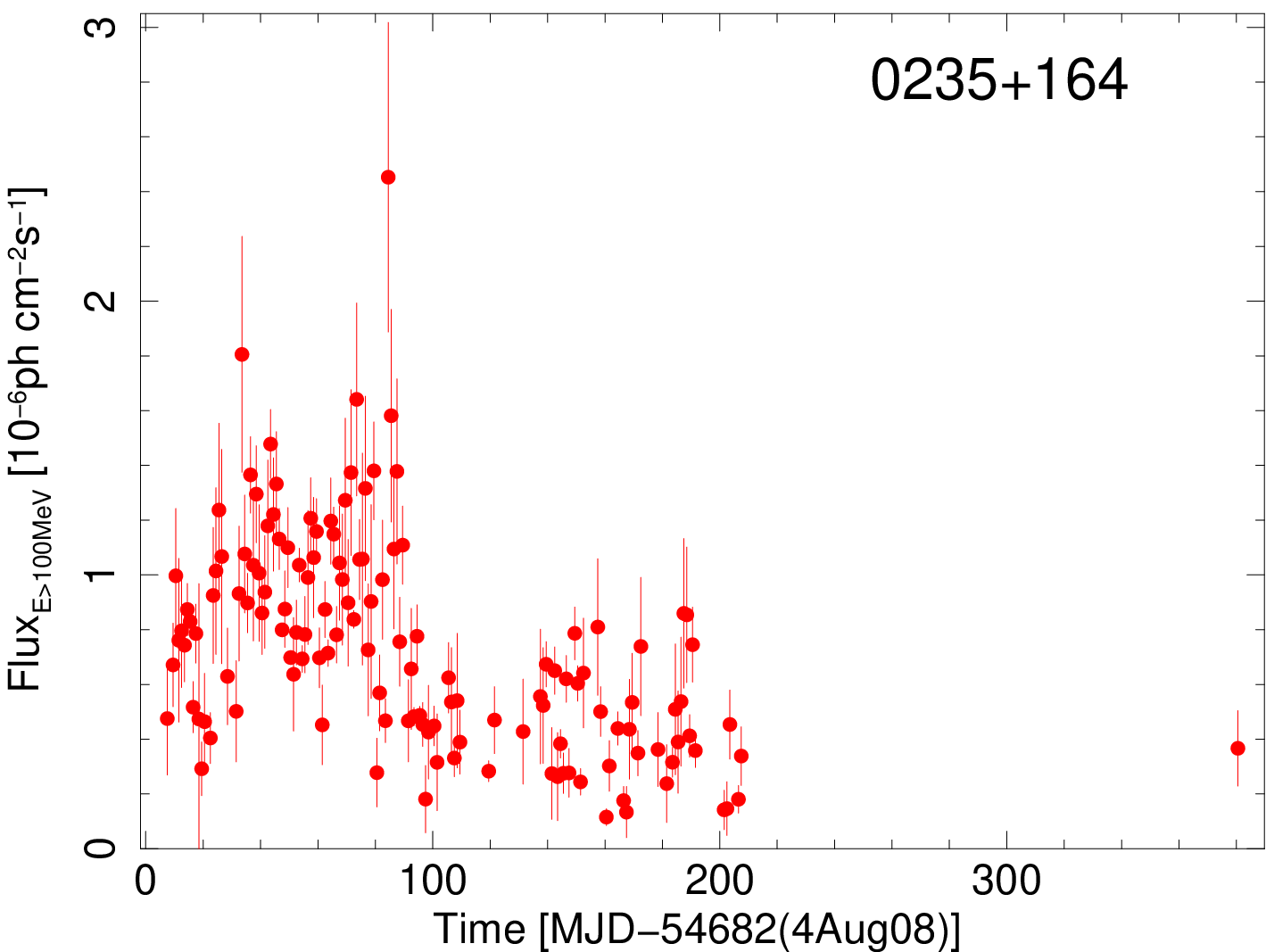}}
\subfigure[]
   {\includegraphics*[angle=270, scale=0.25, clip, trim=0 0 0 0]{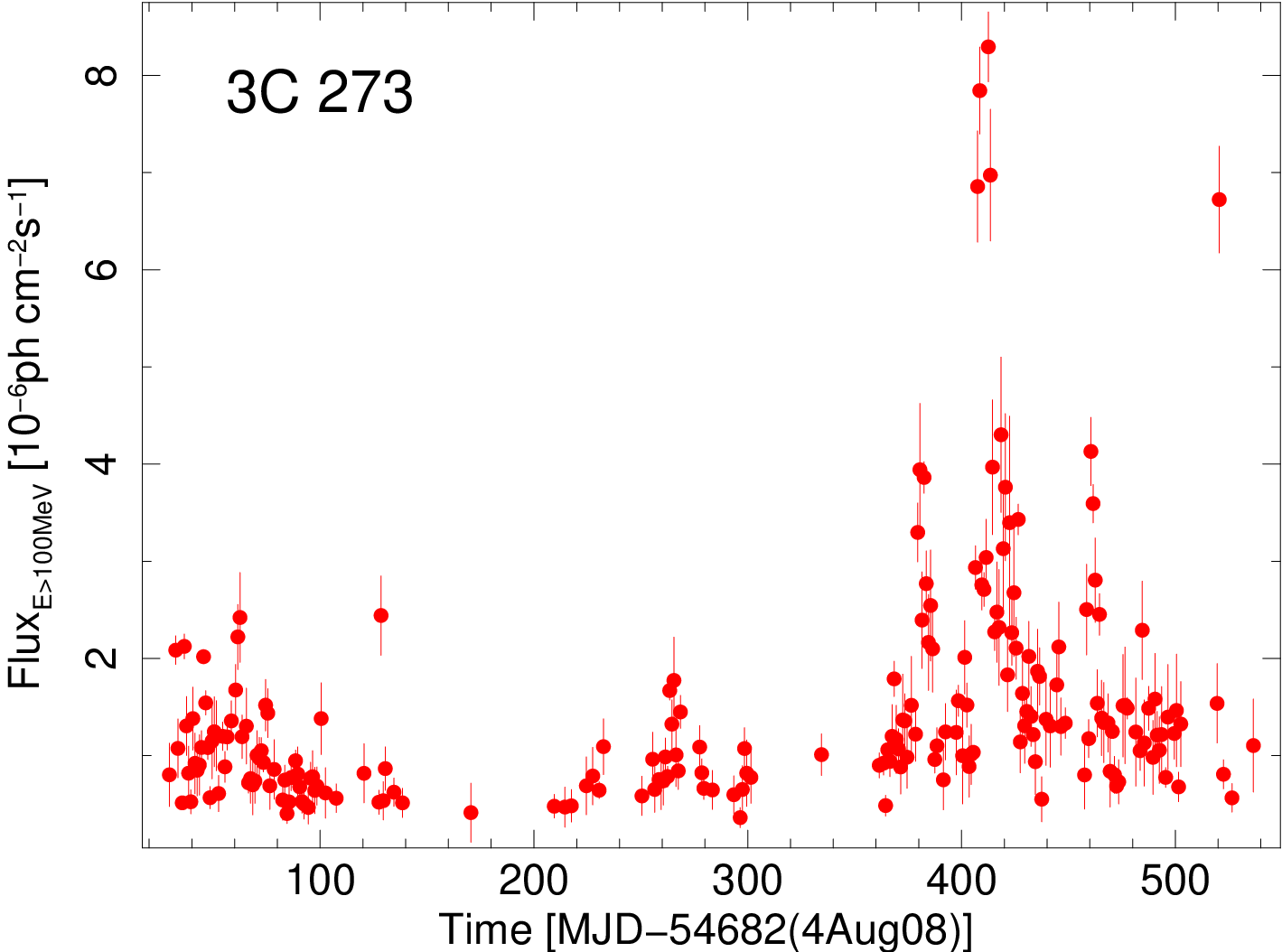}}
\subfigure[]
   {\includegraphics*[angle=270, scale=0.25, clip, trim=0 0 0 0]{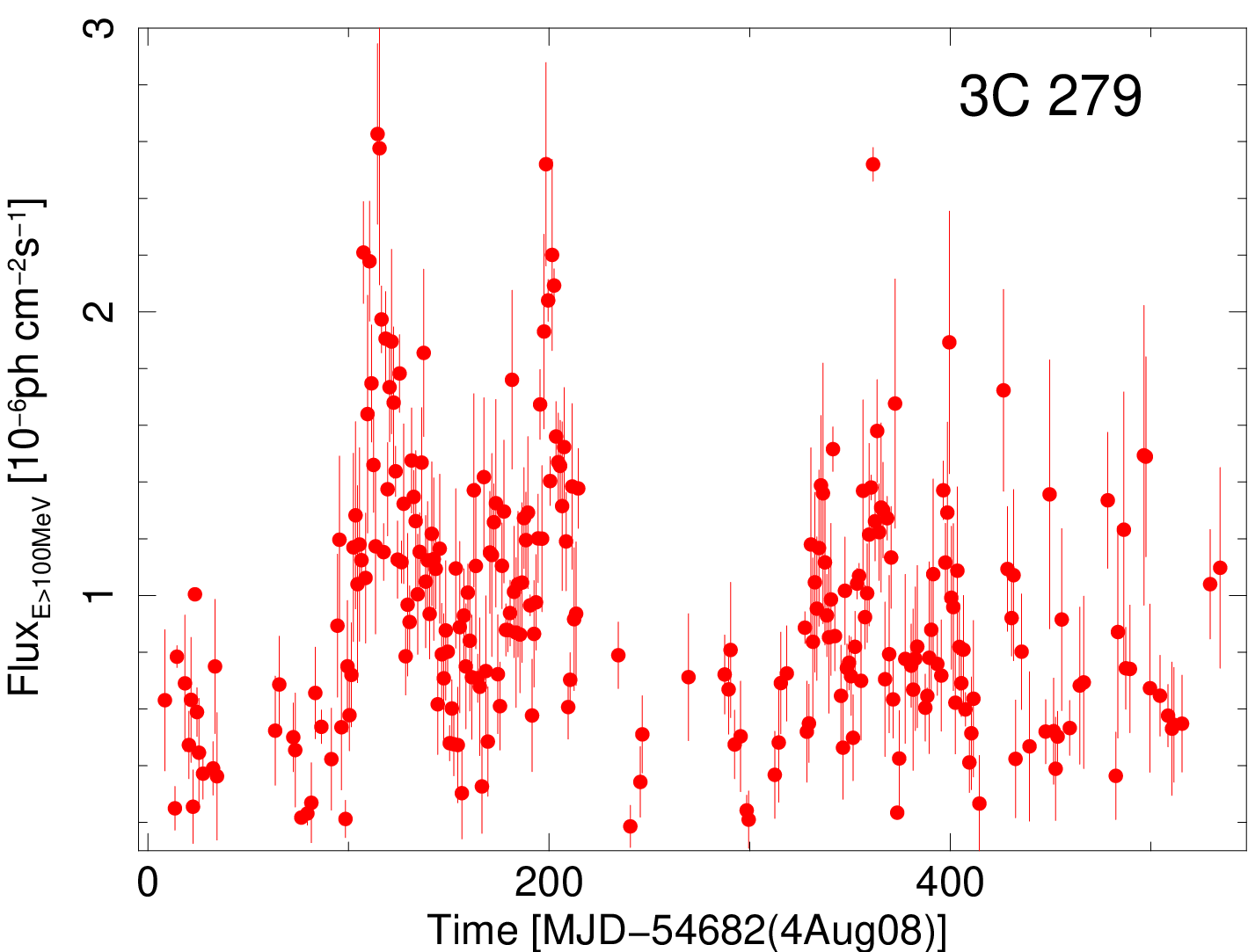}}
\subfigure[]
   {\includegraphics*[angle=270, scale=0.25, clip, trim=0 0 0 0]{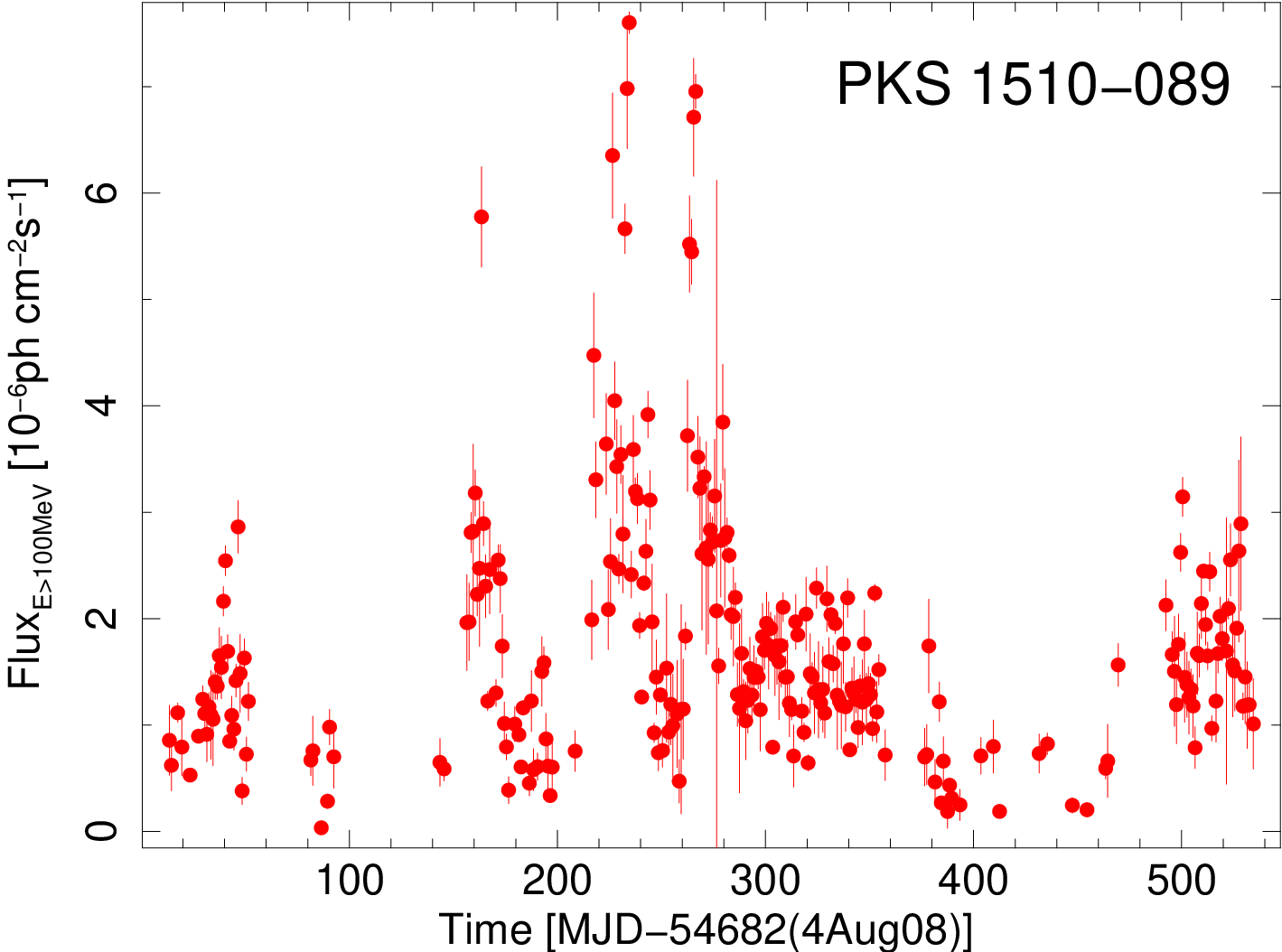}}
\subfigure[]
   {\includegraphics*[angle=270, scale=0.25, clip, trim=0 0 0 0]{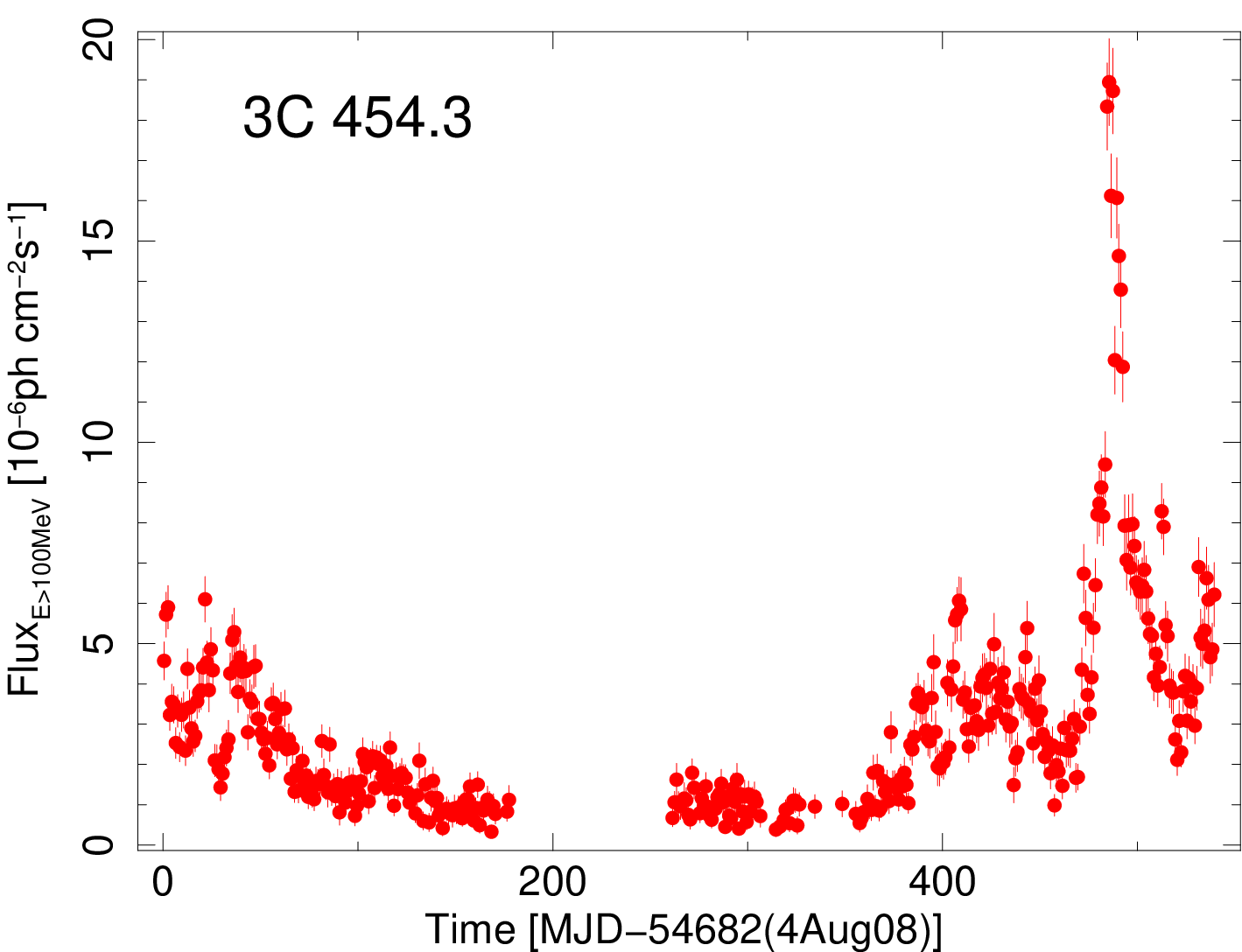}}
\caption{\small Daily $\gamma$-ray light curves ($E>100$ MeV) of the five most variable sources selected for the analysis: (a) 0235+164, (b) 3C~273, (c) 3C~279, (d) 1510-089, (e) 3C~454.3.}
\label{curve}
\end{figure}

Data of the most variable sources have been analysed by means of the public software package \tt LAT Science Tools v.~9.15.2\rm\normalsize, including the calibration files (Instrument Response File, IRF \tt P6\_V3\_DIFFUSE\rm\normalsize), the maps of the spatial diffuse background and the spectrum of the isotropic background, both cosmic and instrumental.
The data sets has been filtered to exclude the events outside the good-time intervals.
The statistical characterization of data sets is based on an unbinned likelihood algorithm, implemented in the \tt gtlike \rm\normalsize task.
This algorithm has been used to extract the flux and the photon index of the source emission,
modeled with a single power-law with the form $F(E) \propto E^{-\Gamma_{\gamma}}$ (where $\Gamma_{\gamma}$ is the photon index).
A gross estimation of the spectral parameters is done with the optimizer \tt DRMNFB\rm\normalsize.
The output of this first run is used as an input for the second optimizer \tt NEWMINUIT\rm\normalsize.
In the following, we will display only statistical errors, since we will not make comparisons with other instruments.
Therefore, we neglected systematic errors, and there is no need of absolute flux calibration.
Performing this analysis, we obtained for the five selected sources the light curves shown in Figure \ref{curve}.

\section{Minimum variability time-scale analysis}

The first step of our analysis is the determination of the minimum variability time-scale,
defined as the shortest doubling  or halving time  among the significant flux variations present in the light curves.
To ensure the variation significancy, we require a difference of $4\sigma$ between the flux values of two adjacents bins, where $\sigma$ is the largest error bar involved in the variation.
Therefore, we calculated the equivalent doubling or halving time for each valid variation, in order to define the shortest ones as $\Delta t_{min}$.
Table \ref{table2} reports the values found for $\Delta t_{min}$, both for the rise and decay phases.
On average, they are very similar.

3C~454.3 shows little variability for most of its light curve,
becoming more active towards the end of the observation period.
In December 2009, 3C~454.3 made an intense outburst, followed by a period of high activity, that reached its climax in April 2010, with a second huge outburst.
The first event starts with the intense activity that can be noticed at the end of our light curve (Figure \ref{curve}e).
These two outbursts are deeply analysed in many works (Ackermann et al.\ 2010, Bonnoli et al.\ 2011, Foschini et al.\ 2011, Tavecchio et al.\ 2010), 
where the minimum variability time-scale obtained is about 3 hours.
More stringent results can be found in literature also for other sources, and the respective time-scales are included in Table \ref{table2}.
These results are obtained through a deeper analysis of high fluxes phases, on which a finer time binning (up to 3 hours) can be performed.
Hence, the results can be more stringent.
We have included in Table \ref{table2} this more restrictive time-scales and the corresponding values Doppler factors ($\delta_{min}$) and blob sizes ($R_{max}$).

The obtained time-scales can be used to derive constraints on some features of the emitting region.
If we assume this region to be a spherical blob in relativistic motion, we can estimate a maximum radius, given by the relation:
\begin{displaymath}
R < c \Delta t_{min} \frac{\delta}{1+z}
\end{displaymath}
where the Doppler factor $\delta$ is derived from the transparency condition to $\gamma$~-rays (Dondi, Ghisellini 1994):
\begin{displaymath}
\delta \geq \bigg[ \frac{\sigma_T d_L^2 (1+z)^{2{\alpha}_x}}{5hc^2} \frac{F(\nu_0)}{\Delta t_{min}} \bigg]^{\frac{1}{4+2{\alpha}_x}}
\end{displaymath}
where $F(\nu_0)=F_{x}(\nu_x/\nu_0)^{{\alpha}_x}$ is calculated for both $\nu_0=1$~GeV and $\nu_0=10$~GeV, in order to take into consideration two different standard frequencies of incident $\gamma$ photons.
The parameter used in this formula for each source are indicated in Table \ref{table1}.
Table \ref{table2} shows the specific values obtained for every source.

We found upper limits for minimum variability time scale to be $\Delta t_{min} \sim1$ day.
These results correspond to emitting regions characterized by maximum sizes in a range from $R \sim 1\cdot10^{-3}\textrm{pc} \sim 3\cdot10^{15}\textrm{cm}$ to $R \sim 9\cdot10^{-3}\textrm{pc} \sim 2.7\cdot10^{16}\textrm{cm}$ and by a minimum Doppler factor in the range $\delta \sim 2-7$,
i.e.\ consistent with results obtained from detailed modeling performed by Ghisellini et al.\ (2010).

\begin{table}
 \centering
\footnotesize
\begin{tabular}{l c c c c c c c c}
\hline
\multicolumn{9}{c}{\textbf{$^{(a)}$Results with $\nu_0=1$~GeV}} \\[1mm]
\textbf{Source} & \multicolumn{4}{c}{\textbf{Rise}} & \multicolumn{4}{c}{\textbf{Decay}}\\
 & $\Delta t_{min}$ & $n_{\sigma}$ & $\delta_{min}$ & $R_{max}$ & $\Delta t_{min}$ & $n_{\sigma}$ & $\delta_{min}$ & $R_{max}$ \\
 & [days] & & & [pc] & [days] & & & [pc] \\ 
\hline
0235+164 & 
$<0.8$ & 4 & 4.3 & $1.6\cdot10^{-3}$ &
$<1.0$ & 6 & 4.1 & $1.9\cdot10^{-3}$ \\
 &
$6$ hr$^{*}$ & & 5.5 & $6.1\cdot10^{-4}$ & 
 & &  &  \\
3C 273 & 
$<0.9$ & 5 & 4.6 & $3.2\cdot10^{-3}$ &
$<0.6$ & 11 & 5.1 & $2.1\cdot10^{-3}$ \\
 &
$5$ hr$^{*}$ & & 6.2 & $9.6\cdot10^{-4}$ & 
 & &  &  \\
3C 279 & 
$<1.5$ & 4 & 3.4 & $2.9\cdot10^{-3}$ &
$<0.5$ & 6 & 4.1 & $1.2\cdot10^{-3}$ \\
1510-089 & 
$<1.1$ & 7 & 2.5 & $1.7\cdot10^{-3}$ &
$<0.9$ & 6 & 2.6 & $1.5\cdot10^{-3}$ \\
 &
$6$ hr$^{*}$ & & 3.4 & $5.4\cdot10^{-4}$ & 
 & &  &  \\
3C 454.3 & 
$<3.5$ & 8 & 4.6 & $7.5\cdot10^{-3}$ &
$<3.0$ & 6 & 4.7 & $6.6\cdot10^{-3}$ \\
 &
$3$~hr$^{*}$ & & 8.5 & $5.0\cdot10^{-4}$ &
 & &  &  \\
\hline
\multicolumn{9}{c}{\textbf{$^{(b)}$Results with $\nu_0=10$~GeV}}\\[1mm]
\textbf{Source} & \multicolumn{4}{c}{\textbf{Rise}} & \multicolumn{4}{c}{\textbf{Decay}}\\
 & $\Delta t_{min}$ & $n_{\sigma}$ & $\delta_{min}$ & $R_{max}$ & $\Delta t_{min}$ & $n_{\sigma}$ & $\delta_{min}$ & $R_{max}$ \\
 & [days] & & & [pc] & [days] & & & [pc] \\ 
\hline
0235+164 & 
$<0.8$ & 4 & 5.3 & $1.9\cdot10^{-3}$ &
$<1.0$ & 6 & 5.1 & $2.3\cdot10^{-3}$ \\
 &
$6$ hr$^{*}$ & & 6.7 & $7.5\cdot10^{-4}$ & 
 & &  &  \\
3C 273 & 
$<0.9$ & 5 & 6.0 & $4.1\cdot10^{-3}$ &
$<0.6$ & 11 & 6.6 & $2.7\cdot10^{-3}$ \\
 &
$5$ hr$^{*}$ & & 7.8 & $1.2\cdot10^{-3}$ & 
 & &  &  \\
3C 279 & 
$<1.5$ & 4 & 4.6 & $3.9\cdot10^{-3}$ &
$<0.5$ & 6 & 5.6 & $1.6\cdot10^{-3}$ \\
1510-089 &  
$<1.1$ & 7 & 3.0 & $2.1\cdot10^{-3}$ &
$<0.9$ & 6 & 3.1 & $1.8\cdot10^{-3}$ \\
 &
$6$ hr$^{*}$ & & 4.1 & $6.5\cdot10^{-4}$ & 
 & &  &  \\
3C 454.3 & 
$<3.5$ & 8 & 6.2 & $1.0\cdot10^{-2}$ &
$<3.0$ & 6 & 6.4 & $8.9\cdot10^{-3}$ \\
 &
$3$~hr$^{*}$ & & 11.5 & $6.7\cdot10^{-4}$ &
 & &  &  \\
\hline
\end{tabular}
\caption{Results of the minimum variability time-scale analysis: $\Delta t_{min}$ is the minimum variability time-scale, from which we derived the minimum Doppler factor $\delta_{min}$ and the maximum size of the emitting region $R_{max}$.
$\delta_{min}$ is obtained from the Dondi and Ghisellini relation (1994) for the jet transparency to $\gamma$-rays, using $\nu_0=1$~GeV$^{(a)}$ and $\nu_0=10$~GeV$^{(b)}$ as standard frequencies of the incident $\gamma$-photons.
$n_{\sigma}$ is the number of sigma of the variation corresponding to the minimum variability time-scale.
$^{*}$Results obtained from several deeper analysis performed during phases of particularly high fluxes of the sources.
0235+164 has been analysed by Foschini et al.\ (2008),
3C~273 by Foschini et al.\ (2011),
1510-089 by Tavecchio et al.\ (2010),
3C~454.3 by Ackermann et al.\ (2010), Bonnoli et al.\ (2011), Foschini et al.\ (2011), Tavecchio et al.\ (2010).}
\label{table2}
\end{table}

\section{Outbursts time-scale analysis}

The second part of our analysis is centered on the interpretation of the outbursts.
A well defined outburst is a substantial variation in flux, lasting at least a few days with a coherent trend in rise and decay.
We found this kind of variation in all the light curves of the selected sources, even if with different shapes and intensities.

\begin{figure}
\centering
\subfigure[]
{\includegraphics*[angle=270, scale=0.3, clip, trim=0 0 0 0]{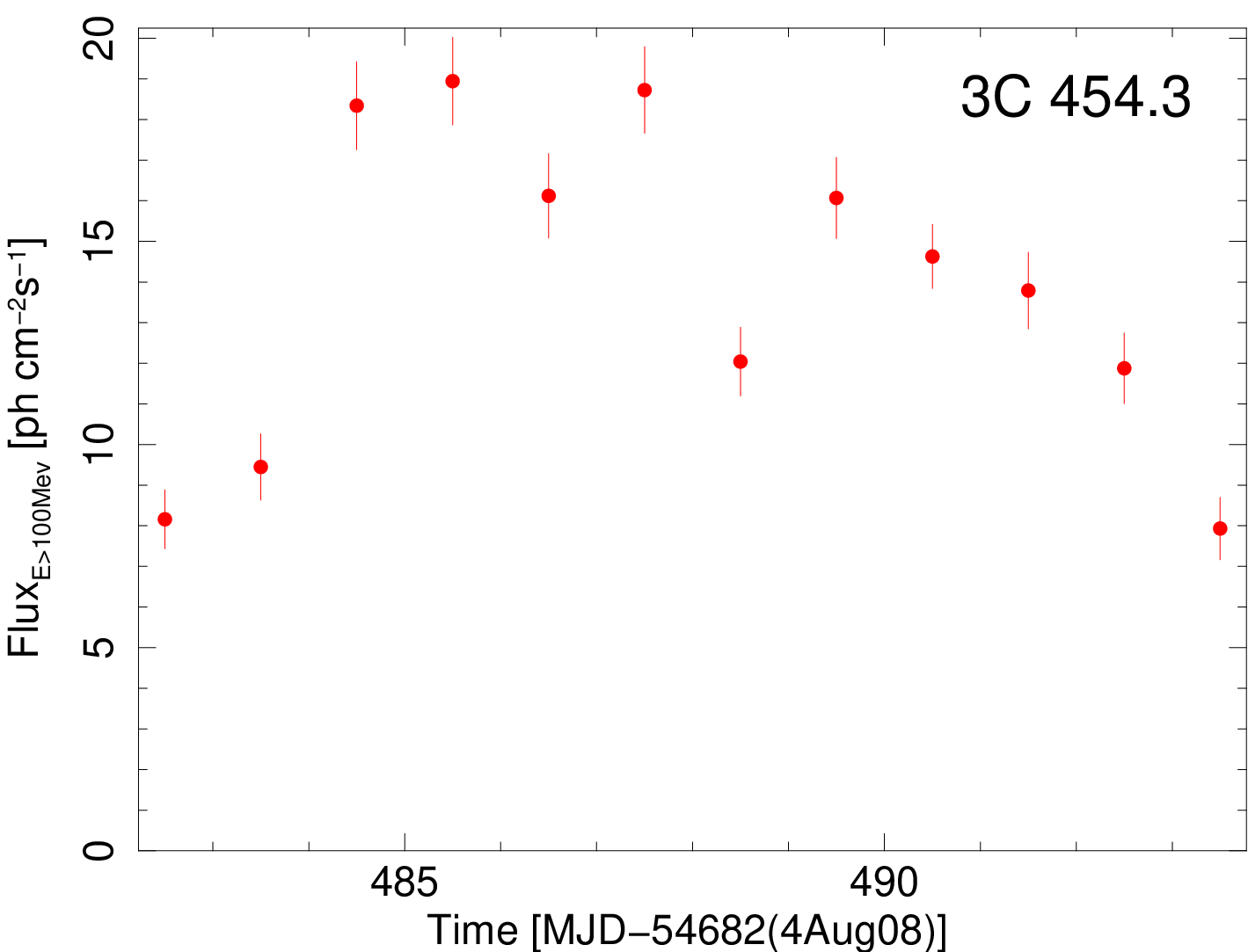}}
\subfigure[]
{\includegraphics*[angle=270, scale=0.3, clip, trim=0 0 0 0]{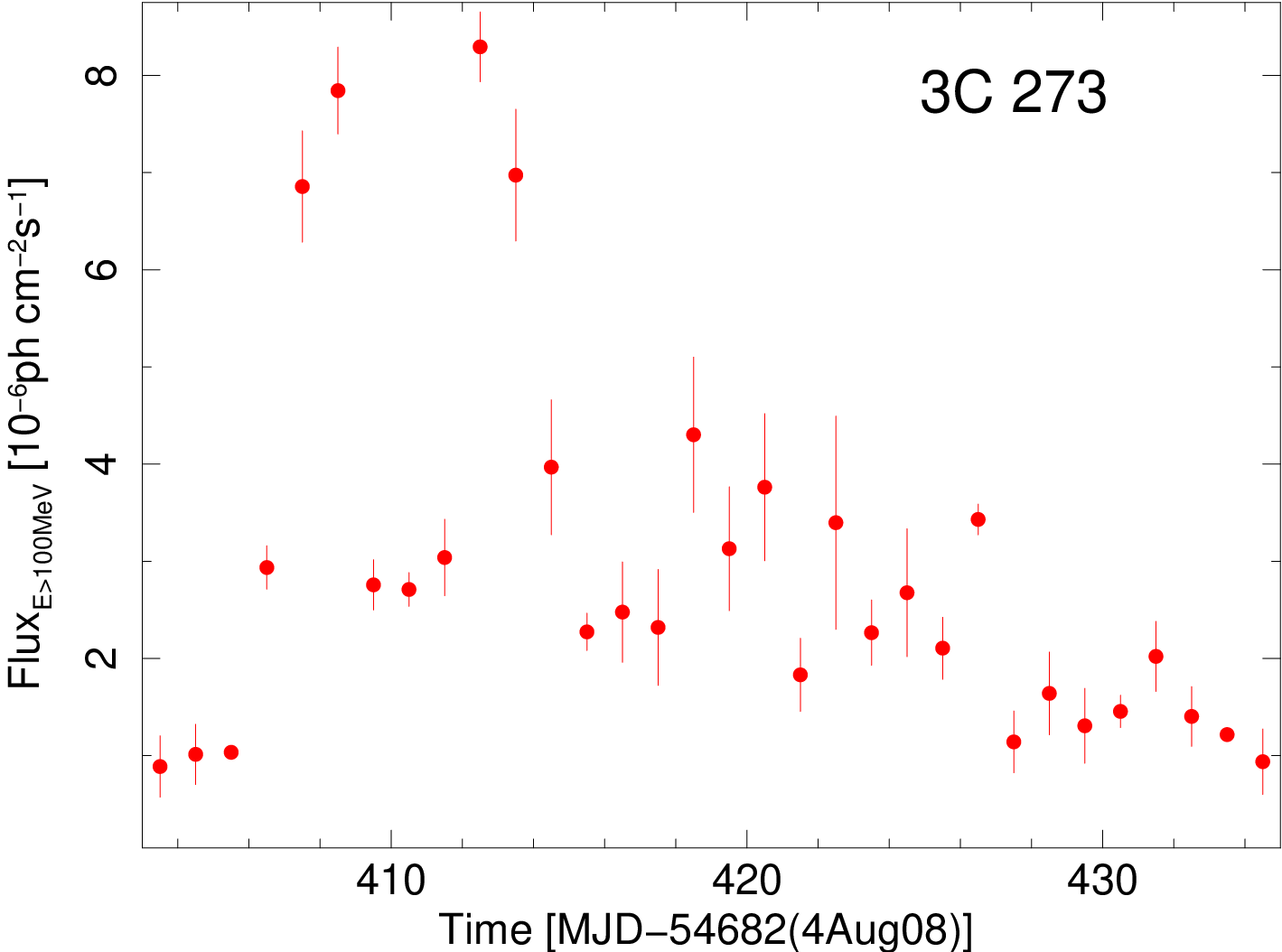}}
\caption{Details of outbursts analysed in 3C~454.3 (a) and 3C~273 (b) $\gamma$-ray light curves.}
\label{zoom}
\end{figure}

The best example of an outburst can be observed during the exceptional activity of 3C~454.3 in December 2009, deeply analysed by Bonnoli et al.\ (2011).
This outburst lasts 11 days and starts from an already high state (figure \ref{zoom}).
In the case of 3C~279, the most coherent case of activity is a couple of long-time variations in the first half of the light curve, as can be seen from the complete light curve in figure \ref{curve}.
The identiﬁcation of a good outburst in this case is made difficult  by the large error bars, related to the
somewhat low flux of this source.
In the central part of the 1510-089 light curve we can see three variations that can fit our definition of outburst, even if less defined than the one in 3C~454.3.
3C~273 presents instead a peculiar profile in its outbursts: every substantial variation seems to be composed of at least two initial spikes, followed by a more gradual decay.
Every spike lasts 3 or 4 days, as can be seen in figure \ref{zoom}.

A first possibility is to ascribe the outburst to the relativistic straight motion of a unique emitting region, assumed to be a spherical blob.
From the analysis of the blob motion feature, we can derive some properties.\\
In a time interval of $\Delta t_e=t_1-t_0$, the blob moves with a velocity $\beta c$ and covers a distance $R=\beta c \Delta t_e$. 
Two photons emitted at the beginning ($t_0$) and at the end ($t_1$) of this path would be separated by a distance $\Delta R = c\Delta t_e - R\cos\theta$,
where $\theta$ is the angle of the blob motion with respect to the photons propagation direction, i.\ e.\ our line of sight.
The two photons are received with a time separation $\Delta t_{outburst}$ given by
\begin{displaymath}
\Delta t_{outburst} = \frac{\Delta R}{c} (1+z) = \Delta t_e (1- \beta\cos\theta) (1+z) \approx \frac{\Delta t_e}{\Gamma^2} (1+z)
\end{displaymath}
where $\Gamma$ is the bulk Lorentz factor.
For blazars, we can assume $\theta \sim \sin\theta \sim 1/\Gamma$ and $\cos\theta \sim \Gamma$, hence $(1-\beta\cos\theta)\approx\Gamma^{-2}$.
The distance $R$ covered by the blob in $\Delta t_{outburst}$ is:
\begin{displaymath}
R = \beta c \Delta t_e = \beta c \Delta t_{outburst} \frac{\Gamma^2}{1+z}
\end{displaymath}
It is likely that during this travel, the blob expands.
To estimate the expansion, we can use the results of the previous section as a reliable approximation of the starting radius $R_{b0}$.
We assume that the blob expansion follows the jet aperture $\psi = 10^{-1} \psi_{-1}$, therefore the blob radius at the end of its motion is:
{\setlength\arraycolsep{2pt}
\begin{eqnarray}
R_{b1} & = & \psi R_{stop} = \psi \Big( \frac{R_{b0}}{\psi} + \beta c \Delta t_{outburst}\frac{\Gamma^2}{1+z}\Big) = {} \nonumber \\
& \approx & \frac{c\Delta t_{min} \Gamma}{1+z} + \psi\beta c \Delta t_{outburst}\frac{\Gamma^2}{1+z} \nonumber
\end{eqnarray}}
where we assumed $\theta\sim1/\Gamma$ and hence $\delta\sim\Gamma$.
Therefore we can estimate the expansion of the blob as:
\begin{displaymath}
\frac{R_{b1}}{R_{b0}} = 
1 + \psi\beta\Gamma\frac{\Delta t_{outburst}}{\Delta t_{min}} \approx 
16 \;\psi_{-1}\;\beta\;\Gamma_{15}\;\Big(\frac{\Delta t_{outburst}/\Delta t_{min}}{10}\Big)
\end{displaymath}
The blob should clearly expand considerably and change quite dramatically its emission properties, contrary to what we observe.

Even if we introduce a smaller jet aperture $\psi\sim1^{\circ}\simeq1.7\cdot10^{-2}\,\psi_{-2}$, of the order of the radio component of the jet measured by VLBI (Jorstad et al.\ 2005), we obtain a significant expansion:
\begin{displaymath}
\frac{R_{b1}}{R_{b0}} \approx 3.6 \; \psi_{-2}\;\beta\;\Gamma_{15}\;\Big(\frac{\Delta t_{outburst}/\Delta t_{min}}{10}\Big)
\end{displaymath}
that corresponds to a variation in the volume by a factor $\sim30$.
Hence, even such a thin aperture angle provokes an alteration large enough to change drastically the blob emitting conditions.
Therefore, such long variations in flux cannot be interpreted with a unique blob displacement.
For this reason, a different interpretation is needed.
One possibility is that the variable luminosity comes from a standing shock.
The emitting plasma flows through the shock at relativistic velocities, but the emitting region remains the same.
Another hypothesis is suggested by the outbursts of 3C~273  (see figure \ref{zoom}).
Indeed, these variations can be interpreted as a superposition of single flares with shorter time-scales.
This model would imply the consecutive formation of blobs and the superposition of their  emission.

\section{Conclusions}

The blazar emitting region can be modeled with a homogeneous spherical blob placed in the jet, with a size comparable to the cross sectional dimension of the jet.
Minimum time-scale variability allows to determine the maximum size of the emitting region.
For the high variability sources that we analysed, the time-scales are found to be $\Delta t_{min}\sim1$ day.
Performing a different pointing with LAT (Foschini et al.\ 2010), or in cases of very strong outbursts, minimum variability could be estimated with more accuracy.
The emitting region derived from  
$\Delta t_{min}$ have maximum sizes that spread from $\sim1\cdot10^{-3}$~pc to $\sim9\cdot10^{-3}$~pc
and lower limits of the bulk Doppler factor in the range $\delta\sim2-7$.
Our minimum variability study is completely model-independent, therefore it can be used as a constraint on models deductions.

Using the outburst time-scale, we tried to describe the relativistic motion of the emitting region.
In the assumed scenario, if the outbursts were produced during a straight motion of the blob along the jet, the emitting region should expand considerably and change dramatically its emission properties, contrary to the  observational evidence.
Therefore, a different hypothesis is needed, such as replacing the formation of a single blob with a more complex mechanism of flux modulation.

\section*{Acknowledgement}

TS firstly acknowledges M.\ Colpi for her many helpful comments and remarks on this work.
This work has made use of data obtained from the High Energy Astrophysics Science Archive Research Center (HEASARC), provided by NASA's Goddard Space Flight Center through the Science Support Center (SSC).
TS acknowledges partial financial support from INAF, IAU and JAXA.

\section*{References}

Abdo A.A., Ackermann M., Ajello M.\ et al.,
Gamma-ray light curves and variability of bright \textit{Fermi}-detected blazars,
ApJ, 722, 520-542, 2010.

Ackermann M., Ajello M., Baldini L.\ et al.,
Fermi Gamma-Ray Space Telescope observations of gamma-ray outbursts from 3C 454.3
in 2009 December and 2010 April,
ApJ, 721, 1383-1396, 2010.

Aharonian F., Akhperjanian A.G., Bazer-Bachi A.R.\ et al.,
An exceptional very high energy gamma-ray flare of PKS 2155-304,
ApJ, 664, L71-L74, 2007.

Aharonian F., Akhperjanian A.G., Anton G.\ et al., 
Simultaneous multiwavelength observations of the second
exceptional $\gamma$-ray flare of PKS2155–304 in July 2006,
A\&A, 502, 749-770, 2009.

Albert J., Aliu E., Anderhub H.\ et al.,
Variable very high energy gamma-ray emission from Markarian 501,
ApJ, 669, 862-883, 2007.

Atwood W.B., Abdo A.A., Ackermann M.\ et al., 
The Large Area Telescope on the Fermi Gamma-Ray Space Telescope Mission,
ApJ, 697, 1071-1102, 2009.

Begelman M.C., Blandford R.D., \& Rees M.J.,
Theory of extragalactic radio sources,
Rev.\ Mod.\ Ph., 56, 255-351, 1984.

Begelman M.C., Fabian A.C., \& Rees M.J., 
Implications of very rapid TeV variability in blazars,
MNRAS, 384, L19-L23, 2008.

Bonnoli G., Ghisellini G., Foschini L.\ et al., 
The $\gamma$-ray brightest days of the blazar 3C 454.3,
MNRAS, 410, 368-380, 2011.

Dondi L., \& Ghisellini G., 
Gamma-ray-loud blazars and beaming,
MNRAS, 273, 583-595, 1995.

Foschini L., Ghisellini G., Tavecchio F.\ et al., 
X.ray/UV/optical follow-up of the blazar PKS 2155-304 after the giant TeV flares of 2006 July,
ApJ, 657, L81-L84, 2007.

Foschini L., Longo F.\ \& Iafrate G.,
Fermi/LAT detection of strong activity on short timescales of the blazar AO 0235+164,
ATel 1784, 2008.

Foschini L., Tagliaferri G., Ghisellini G.\ et al., 
Does the gamma-ray flux of the blazar 3C 454.3 vary on sub-hour time scales?
MNRAS, 408, 448-451, 2010.

Foschini L., Ghisellini G., Tavecchio F.\ et al.,
Search for the shortest variability at gamma rays in flat-spectrum radio quasars,
submitted [arXiv: 1101.1085v1], 2011.

Ghisellini G., Tavecchio F., Foschini L.\ et al.,
General physical properties of bright Fermi blazars,
MNRAS, 402, 497-518, 2010.

Jorstad S.\ G., Marscher A.\ P., Lister M.\ L.\ et al.,
Polarimetric observations of 15 Active Galactic Nuclei at high frequencies:
jet kinematics from bimonthly monitoring with the Very Long Baseline Array.
AJ, 130, 1418-1465, 2005.

Mattox J.R., Bertsch D.L., Chiang J.\ et al., 
The Likelihood Analysis of EGRET Data,
ApJ, 461, 396-407, 1996.

Mattox J.R., Wagner S.J., Malkan M.\ et al., 
An intense gamma-ray flare of PKS 1622-297,
ApJ, 476, 692-697, 1997.

Tavecchio F., Ghisellini G., Bonnoli G., \& Ghirlanda G., 
Constraining the location of the emitting region in Fermi blazars through rapid $\gamma$-ray variability,
MNRAS, 405, L94-L98, 2010.

Ulrich M.H., Maraschi L., \& Urry C.M., 
Variability of Active Galactic Nuclei,
ARA\&A, 35, 445-502, 1997.

Wehrle A.E., Pian E., Urry C.M.\ et al., 
Multiwavelenght observations of a dramatic high-energy flare in the blazar 3C 279,
ApJ, 497, 178-187, 1998.


\clearpage

\end{document}